\documentclass[twocolumn,floatfix]{aastex6}

\usepackage{graphicx}
\usepackage{dcolumn}
\usepackage{bm}
\usepackage{latexsym}
\usepackage{amsmath}
\usepackage{amssymb}

\newcommand{\bfo}{\overline\beta_{\rm f}}
\newcommand{\gfo}{\overline\gamma_{\rm f}}
\newcommand{\dXkp}{\delta X_{\boldsymbol{k_\perp}}}
\newcommand{\dVkp}{\delta V_{\boldsymbol{k_\perp}}}
\newcommand{\bsk}{\boldsymbol{k}}
\newcommand{\bskp}{\boldsymbol{k_\perp}}

\newcommand{\BibitemShut}[1]{}

\begin{document}


\title{Corrugation of relativistic magnetized shock waves}


\author{Martin Lemoine} 
\author{Oscar Ramos} \affiliation{Institut
d'Astrophysique de Paris, \\ UMR 7095-CNRS, Universit\'e Pierre et Marie
Curie, \\ 98bis boulevard Arago, F-75014 Paris, France}
\and
\author{Laurent Gremillet} \affiliation{CEA, DAM, DIF \\ F-91297 Arpajon cedex, France}



\begin{abstract}
  As a shock front interacts with turbulence, it develops corrugation
  which induces outgoing wave modes in the downstream plasma. For a
  fast shock wave, the incoming wave modes can either be fast
  magnetosonic waves originating from downstream, outrunning the
  shock, or eigenmodes of the upstream plasma drifting through the
  shock. Using linear perturbation theory in relativistic MHD, this
  paper provides a general analysis of the corrugation of relativistic
  magnetized fast shock waves resulting from their interaction with
  small amplitude disturbances. Transfer functions characterizing the
  linear response for each of the outgoing modes are calculated as a
  function of the magnetization of the upstream medium and as a
  function of the nature of the incoming wave. Interestingly, if the
  latter is an eigenmode of the upstream plasma, we find that there
  exists a resonance at which the (linear) response of the shock
  becomes large or even diverges. This result may have profound
  consequences on the phenomenology of astrophysical relativistic
  magnetized shock waves.
\end{abstract}




\section{Introduction} 
\label{sec:introduction}
The physics of relativistic shock waves, in which the unshocked plasma
enters the shock front with a relative relativistic velocity $v_{\rm
  sh}\,\sim\,c$, is a topic which has received increased attention
since the discovery of various astrophysical sources endowed with
relativistic outflows, such as radio-galaxies, micro-quasars, pulsar
wind nebulae or gamma-ray bursts. In those objects, the relativistic
shock waves are believed to play a crucial role in the dissipation of
plasma bulk energy into non-thermal particle energy, which is then
channeled into non-thermal electromagnetic radiation (or possibly,
high energy neutrinos and cosmic rays). The various manifestations of
these high energy sources have been a key motivation to understand the
physics of collisionless shock waves and of the ensuing particle
acceleration
processes~\citep[e.g.][]{2011A&ARv..19...42B,2012SSRv..173..309B,2015SSRv..191..519S}
for reviews. The nature of the turbulence excited in the vicinity of
these collisionless shocks remains a nagging open question, which is
however central to all the above topics, since it directly governs the
physics of acceleration and, possibly, radiation.

The physics of shock waves in the collisionless regime has itself been
a long-standing problem in plasma physics, going back to the
pioneering studies of ~\citet{1963JNuE....5...43M}, with intense
renewed interest related to the possibility of reproducing such shocks
in laboratory
astrophysics~\citep[e.g.][]{2011PhRvL.106q5002K,2012ApJ...749..171D,Park201238,Huntington_NP_11_173_2015,Park_PHP_22_056311_2015}. The
generation of relativistic collisionless shock waves is also already
envisaged with future generations of
lasers~\citep[e.g.][]{Chen_PRL_114_215001_2015, 2015PhRvL.115u5003L}.

One topic of general interest, with direct application to the above
fields, is the stability of shock waves. The study of the corrugation
instability of a shock wave goes back to the early works of
~\citet{1958JETP....6..739D} and \citet{1958JETP....6.1179K}, see also
~\citet{1982SvAL....8..320B}, \citet{LandauLifshitz87} or more
recently ~\citet{2000PhRvL..84.1180B}. General theorems assuming
polytropic equations of state ensure the stability of shock waves
against corrugation, in the relativistic~\citep{1986PhFl...29.2847A}
and/or magnetized
regime~\citep{1964PhFl....7..700G,1967PhFl...10..782L,McKW71},
although instability may exist in other
regimes~\citep[e.g.][]{TCST97}. In any case, the stability against
corrugation does not preclude the possibility of spontaneous emission
of waves by the shock front, as discussed in the above references.

The interaction of the shock front with disturbances thus represents a
topic of prime interest, as it may lead to the corrugation of the
shock front and to the generation of turbulence behind the shock, with
possibly large amplification. The transmission of upstream Alfv\'en
waves through a sub-relativistic shock front has been addressed, in
particular, by~\citet{1986MNRAS.218..551A}; more recently,
~\citet{2012MNRAS.422.3118L} has reported on numerical MHD simulations
of the interaction of a fast magnetosonic wave impinging on the
downstream side of a relativistic shock front.

The present paper proposes a general investigation of the corrugation
of relativistic magnetized collisionless shock waves induced by either
upstream or downstream small amplitude perturbations. This study is
carried out analytically for a planar shock front in linearized
relativistic MHD.  This problem is addressed as
follows. Section~\ref{sec:gen} provides some notations as well as the
shock crossing conditions to the first order in perturbations, which
relate the amplitude of shock corrugation to the amplitude of incoming
and outgoing MHD perturbations of the flow. Section~\ref{sec:dscatt}
is devoted to the interaction of a fast magnetosonic wave originating
from downstream and to its scattering off the shock front, with
resulting outgoing waves and shock
corrugation. Section~\ref{sec:utrans} discusses the transmission of
upstream entropy and Alfv\'en perturbations into downstream
turbulence. It reveals, in particular, that there exist resonant
wavenumbers of the turbulence for which the amplification of the
incoming wave, and consequently the amplitude of the shock crossing,
becomes formally infinite. This resonant excitation of the shock front
by incoming upstream turbulence may have profound implications for our
understanding of astrophysical shock waves and the associated
acceleration processes.

\section{General considerations}\label{sec:gen}
We assume here a configuration in the rest frame of the downstream
(shocked) plasma in which the magnetic field is exactly perpendicular
to the shock normal, and in which the upstream is inflowing into the
shock along the shock normal. The former assumption is a very good
approximation at relativistic shock waves~\citep{1990ApJ...353...66B}
because of Lorentz boost effects, which enhance the in-plane
components of the magnetic field by the relative Lorentz factor
between the upstream and the downstream plasma, notwithstanding the
further compression resulting from the jump at the shock. The latter
is an assumption which allows to keep the problem tractable; shock
crossing at an oblique shock can be obtained analytically but at the
price of an implicit
equation~\citep{1987PhFl...30.3045M,1999JPhG...25R.163K}, which renders
a further perturbative treatment quite complex.

\subsection{Steady planar normal shock}\label{sec:appS}
Although the equations of shock crossing and their solutions are known
for a steady planar normal shock, it is useful to recall them in order
to specify the present notations. At a shock surface defined by its
normal four-vector $\ell_\mu$, these shock crossing conditions are
expressed as:
\begin{eqnarray}
\left[nu^\mu\,\ell_\mu\right]&\,=\,&0,\nonumber\\
\left[T^{\mu\nu}\,\ell_\mu\right]&\,=\,&0,\nonumber\\
\left[^{\star}F^{\mu\nu}\,\ell_\mu\right]&\,=\,&0,\label{eq:sc0}
\end{eqnarray}
with, in the ideal MHD description:
\begin{eqnarray}
T^{\mu\nu}&\,=\,&\left(w + \frac{b_\alpha b^\alpha}{4\pi}\right)u^\mu u^\nu\,+\,
\left(p+\frac{b_\alpha b^\alpha}{8\pi}\right)\eta^{\mu\nu}-\frac{b^\mu b^\nu}{4\pi},\nonumber\\
^{\star}F^{\mu\nu}&\,=\,&\frac{1}{2}\epsilon^{\mu\nu\alpha\beta}F_{\alpha\beta},
\end{eqnarray}
and the following definitions: $F^{\mu\nu}$ denotes the usual
electromagnetic strength tensor, the metric $\eta^{\mu\nu}$ has
signature $(-,+,+,+)$, $\epsilon^{\mu\nu\alpha\beta}$ denotes the
Levi-Civita tensor ($+1$ for an even permutation of the indices),
while $w\,\equiv\,e+p$ represents the fluid enthalpy, and $e,p, n$
respectively correspond to the fluid energy density, pressure and
density; finally, $u^\mu\,=\,\left(\gamma,\boldsymbol{u}\right)$
represents the fluid four-velocity (we use natural units with $c=1$
everywhere in this paper) and $b^\mu$ the magnetic four-vector:
\begin{equation}
b^\mu\,=\,\left[u^iB_i,\left(\boldsymbol{B}+u^iB_i\,\boldsymbol{u}\right)/u^0\right]
\end{equation}
written in terms of the (apparent) magnetic field vector
$\boldsymbol{B}$. Finally, in the MHD description, one has:
\begin{equation}
^{\star}F^{\mu\nu}\,=\,u^\mu b^\nu-u^\nu b^\mu
\end{equation}

The shock crossing conditions are then most conveniently expressed in
the downstream rest frame, in which the shock surface is described by
\begin{equation}
\overline\Phi(x)\,=\,x-\bfo t\,=\,0
\end{equation}
with corresponding shock normal:
\begin{eqnarray}
\overline\ell_\mu&\,=\,&\frac{\partial_\mu\overline\Phi}{\left\vert\partial_\alpha\overline\Phi\partial^\alpha\overline\Phi\right\vert^{1/2}}\nonumber\\ &\,=\,&
\left(-\gfo\bfo,\gfo,0,0\right)\label{eq:sn0}
\end{eqnarray}
where $\gfo\,\equiv\,\left(1-\bfo^2\right)^{-1/2}$ denotes the bulk
Lorentz factor of the shock front relative to downstream. 

\begin{figure}[ht]
\includegraphics[width=0.8\columnwidth]{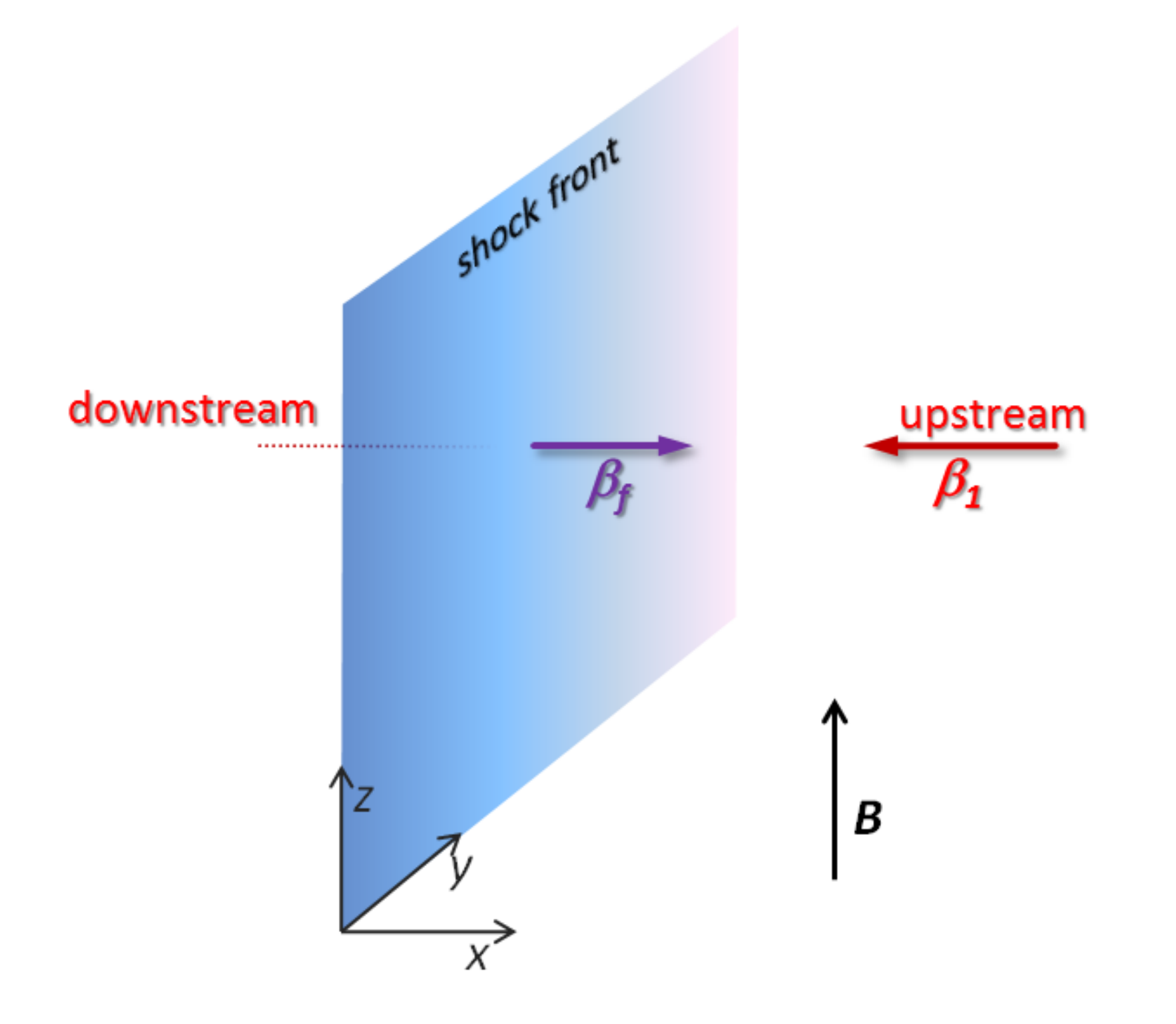}
\caption{Sketch of the shock configuration: in the downstream plasma
  rest frame, the shock front moves at velocity $\bfo\,>\,0$ while the
  upstream (unshocked) plasma is inflowing at velocity
  $\beta_1\,\simeq\,-1$. As indicated, the magnetic field carried by
  the inflowing upstream plasma lies along
  $\boldsymbol{z}$.\label{fig:skesh}}
\end{figure}

Henceforth, downstream quantities are indexed with $_2$ while upstream
quantities are indexed with $_1$; the notation
$b_{1,2}\,\equiv\,B_{1,2}/\gamma_{1,2}$ is also used in the equations
below. In the downstream frame, one has $u_2^\mu\,=\,(1,0,0,0)$ and
$u_1^\mu\,=\,(\gamma_1,\gamma_1\beta_1,0,0)$. We also use the
short-hand notations for the generalized enthalpy and pressure:
\begin{equation}
W\,\equiv\,w+\frac{b_\alpha b^\alpha}{4\pi},\quad
P\,\equiv\,p+\frac{b_\alpha b^\alpha}{8\pi},
\end{equation}
Figure~\ref{fig:skesh} shows a sketch of the planar shock
configuration, emphasizing the notions of the shock velocity ($\bfo$)
and of the upstream plasma ($\beta_1$), both expressed relative to the
downstream frame.

Given Eq.~(\ref{eq:sn0}), the shock crossing equations~(\ref{eq:sc0})
break down to:
\begin{eqnarray} 
n_1\gamma_1\left(\beta_1-\bfo\right)&\,=\,&-n_2\bfo\\
b_1\gamma_1\left(\beta_1-\bfo\right)&\,=\,&-b_2\bfo\\
W_1\gamma_1^2\left(\beta_1-\bfo\right) + \bfo P_1&\,=\,&
-\bfo\left(W_2 -P_2\right)\\
W_1\gamma_1^2\beta_1\left(\beta_1-\bfo\right) + P_1&\,=\,&
P_2
\end{eqnarray}
These equations are easily solved in the strong shock
($P_1\,\ll\,P_2$) and relativistic limit
($\gamma_1\,\gg\,1$). Defining the magnetization parameter:
\begin{equation}
\sigma_1\,\equiv\,\frac{b_1^2}{4\pi w_1},
\end{equation}
so that $W_1\,=\,b_1^2\left(1+\sigma_1^{-1}\right)$, assuming a cold
incoming plasma $w_1\,=\,n_1m$, one finds
\begin{equation}
\bfo\,\simeq\,\frac{2\sigma_1+1+\left[1+16\sigma_1\left(1+\sigma_1\right)\right]^{1/2}}{6(1+\sigma_1)}
\end{equation}
up to corrections of order $1/\gamma_1^2$. This solution matches the
standard result of \citet{KC84a}, although it is given here in a
simpler form. It can be approximated as:
\begin{equation}
\bfo\,\simeq\,\begin{cases}\displaystyle{\frac{1}{3}+\frac{4\sigma_1}{3}} & \left(\sigma_1\,\ll\,1\right)\\ & \\ \displaystyle{1-\frac{1}{2\sigma_1}} &\left(\sigma_1\,\gg\,1\right)\end{cases}
\end{equation}
Once $\bfo$ is known, the shock crossing conditions
immediately give $n_2$ and $b_2$ as a function of $n_1$ and $b_1$
respectively. One also derives
\begin{eqnarray}
\frac{b_2^2}{4\pi
  w_1}&\,\simeq\,&\frac{1}{\bfo}\left(1+\sigma_1\right)\gamma_1^2\left(\beta_1-\bfo\right)\left(3\beta_1\bfo+1\right),\\ 
\frac{w_2}{w_1}&\,\simeq\,&
-\frac{2}{\bfo}\left(1+\sigma_1\right)\gamma_1^2\left(\beta_1-\bfo\right)\left(\beta_1\bfo+1\right)
\end{eqnarray}
so that, for instance,
\begin{equation}
\sigma_2\,=\,\frac{b_2^2}{4\pi w_2}\,\simeq\,\begin{cases}
3\sigma_1 &(\sigma_1\,\ll\,1)\\ 2\sigma_1 & (\sigma_1\,\gg\,1)\end{cases}
\end{equation}

\subsection{Shock corrugation}
We now consider the influence of perturbations in the background flow
on the shock.  As the perturbations impinge on the shock, they induce
a deformation of the shock surface, which can be written up to first
order in the perturbations as:
\begin{eqnarray}
\Phi(x)&\,=\,&\overline\Phi(x,t) + \delta\Phi\left(\boldsymbol{x_\perp},t\right)\nonumber\\
&\,=\,& x-\bfo t - \delta X\left(\boldsymbol{x_\perp},t\right)
\end{eqnarray}
with $\boldsymbol{x_\perp}\,\equiv\,(y,z)$.  Corrrespondingly, the
normal of the perturbed shock surface is written to first order in the
perturbations as:
\begin{equation}
\ell_\mu\,=\,\overline\ell_\mu + \delta\ell_\mu
\end{equation}
with
\begin{equation}
\delta\ell_\mu\,=\,-\frac{\partial_\mu\delta X}{\left\vert
\partial_\alpha\overline\Phi\partial^\alpha\overline\Phi\right\vert^{1/2}}
+ \frac{\partial_\mu\overline\Phi\,\partial_\beta\delta X\partial^\beta\overline\Phi}{\left\vert
\partial_\alpha\overline\Phi\partial^\alpha\overline\Phi\right\vert^{3/2}}
\end{equation}

For the purpose of this Section, consider a harmonic perturbation on a
single scale $\bskp$: 
\begin{equation}
\delta X\left(\boldsymbol{x_\perp},t\right)\,=\,
\dXkp(t)\,\, e^{i\boldsymbol{k_\perp}\cdot\boldsymbol{x_\perp}}
\end{equation}
with a similar decomposition for all other variables.  One then
obtains:
\begin{eqnarray}
\delta\ell_{\bskp,\mu}&\,=\,&\biggl(-\gfo^3\,\dVkp,\,\,
\gfo^3\bfo\,\dVkp,\,\, -ik_y\,\gfo\dXkp,\nonumber\\
&&\quad\quad -ik_z\,\gfo\dXkp\biggr)
\end{eqnarray}
with
\begin{equation}
\dVkp\,\equiv\,\frac{{\rm d}}{{\rm d}t}\dXkp
\end{equation}

The deformation of the shock surface then implies the existence of
fluctuations of the various quantities of the downstream plasma at the
shock. These quantities are formally obtained by the solution to the
shock crossing conditions at the first order in the perturbations:
\begin{eqnarray}
\left[\ell_\mu \,\delta\left(n u^\mu\right) + \delta \ell_\mu n u^\mu\right]&\,=\,&0\\
\left[\ell_\mu\delta T^{\mu\nu} + \delta\ell_\mu T^{\mu\nu}\right]&\,=\,&0\\
\left[\ell_\mu\,\delta^{\star}F^{\mu\nu}+\delta\ell_\mu\,^{\star}F^{\mu\nu}\right]&\,=\,&0
\end{eqnarray}

\begin{widetext}

For harmonic perturbations on the shock front plane in transverse
Fourier space, the above equations can be written as follows:
\begin{eqnarray}
n_2\delta\beta_{2x,\bskp}
-\bfo \delta n_{2,\bskp}+ \gfo^2\left[\gamma_1n_1\left(1-\beta_1\bfo\right)-n_2\right]\dVkp &\,=\,& R_1\nonumber\\
 - \bfo \delta n_{2,\bskp}-\bfo \frac{B_2}{4\pi} \delta B_{2z,\bskp}+\frac{\bfo}{1-\hat\gamma_2}\delta p_{2,\bskp}
+ W_2 \delta\beta_{2x,\bskp}+ \gfo^2\left[\gamma_1^2W_1\left(1-\beta_1\bfo\right)-W_2+P_2-P_1\right]
\dVkp&\,=\,&R_2\nonumber\\
 \delta p_{2,\bskp} + \frac{B_2}{4\pi}\delta B_{2z,\bskp} - W_2 \bfo\delta\beta_{2x,\bskp}+\gfo^2\left[\gamma_1^2W_1\beta_1\left(1-\beta_1\bfo\right)+\bfo\left(P_2-P_1\right)\right]\dVkp &\,=\,&R_3\nonumber\\
-W_2\bfo\delta\beta_{2y,\bskp}-ik_y\left(P_2-P_1\right)\dXkp&\,=\,&R_4\nonumber\\
-w_2\bfo\delta\beta_{2z,\bskp}-\frac{B_2}{4\pi}\delta B_{2x,\bskp}+ik_z\left[\frac{B_2^2}{4\pi}-\frac{B_1^2}{4\pi\gamma_1^2}
-\left(P_2-P_1\right)\right]\dXkp&\,=\,&R_5\nonumber\\
-\delta B_{2x,\bskp} +ik_z\left(B_2-B_1\right)\dXkp&\,=\,&R_6\nonumber\\
-\bfo\delta B_{2x,\bskp} -ik_z\beta_1B_1\dXkp&\,=\,&R_7\nonumber\\
-\bfo \delta B_{2y,\bskp}&\,=\,&R_8\nonumber\\
 B_2\delta\beta_{2x,\bskp} - \bfo \delta B_{2z,\bskp} - \gfo^2\left[B_2-B_1\left(1-\beta_1\bfo\right)\right]\dVkp&\,=\,&R_9\nonumber\\
&&
\label{eq:sysbound}
\end{eqnarray}
where the quantities $R_i$ are expressed in terms of perturbations of
the upstream plasma; they thus all vanish if the upstream plasma is
unperturbed and the shock is corrugated by downstream magnetosonic
waves, as discussed in Sec.~\ref{sec:dscatt}.

One finds:
\begin{eqnarray}
R_1&\,=\,& (\beta_1-\bfo)\gamma_1\delta n_{1,\bskp} +
\gamma_1^3n_1\left(1-\beta_1\bfo\right)\delta\beta_{1x,\bskp}\nonumber\\ 
R_2&\,=\,&
\gamma_1^2\left(\beta_1-\bfo\right)\delta n_{1,\bskp}+
\frac{B_1}{4\pi}\left[\beta_1\left(1-\beta_1\bfo\right)+\beta_1-\bfo\right]\delta B_{1z,\bskp} +
W_1\gamma_1^4\left[1-\beta_1\bfo+\beta_1\left(\beta_1-\bfo\right)\right]
\delta\beta_{1x,\bskp}-\nonumber\\
&& \quad \frac{B_1^2}{4\pi}\beta_1\left[\bfo+
2\left(\beta_1-\bfo\right)\gamma_1^2\right]\delta\beta_{1x,\bskp}
\nonumber\\ 
R_3&\,=\,&\gamma_1^2\beta_1\left(\beta_1-\bfo\right)\delta n_{1,\bskp}+
\frac{B_1}{4\pi\gamma_1^2}\left[1+2\beta_1\left(\beta_1-\bfo\right)\gamma_1^2\right]\delta
B_{1z,\bskp} +\gamma_1^4W_1\left[\beta_1-\bfo+\beta_1\left(1-\beta_1\bfo\right)\right]
\delta\beta_{1x,\bskp} -\nonumber\\
&&  \quad \frac{B_1^2}{4\pi}\beta_1\left[1+2\beta_1\left(\beta_1-\bfo\right)\gamma_1^2\right]
\delta\beta_{1x,\bskp}\nonumber\\ 
R_4&\,=\,&W_1\gamma_1^2\left(\beta_1-\bfo\right)\delta\beta_{1y,\bskp}\nonumber\\ 
R_5&\,=\,&-\frac{B_1}{4\pi\gamma_1^2}\left[1+\beta_1\left(\beta_1-\bfo\right)\gamma_1^2\right]\delta B_{1x,\bskp} +w_1\gamma_1^2\left(\beta_1-\bfo\right)\delta\beta_{1z,\bskp}\nonumber\\
R_6&\,=\,&-\delta B_{1x,\bskp}\nonumber\\
R_7&\,=\,&-\bfo \delta B_{1x,\bskp}\nonumber\\
R_8&\,=\,&\left(\beta_1-\bfo\right)\delta B_{1y,\bskp}\nonumber\\
R_9&\,=\,&\left(\beta_1-\bfo\right)\delta B_{1z,\bskp}+B_1\delta \beta_{1x,\bskp}
\label{eq:sysrhs}
\end{eqnarray}
\end{widetext}
Note that the above equations are valid to first order in the
perturbations, but they apply equally well for relativistic and
non-relativistic shocks, as for magnetized and unmagnetized plasmas.

These equations can be simplified through the use of the unperturbed
shock crossing conditions. In particular, one easily notices that in
the system Eq.~(\ref{eq:sysbound}), the sixth and seventh equations
are redundant, hence this system contains only eight independent
equations. Nevertheless, this suffices to determine the eight
perturbations of the MHD fluid in terms of the quantities
determining the degree of corrugation, i.e. $\dXkp$ and $\dVkp$. In
this sense, the problem is well-posed.

The above equations can be solved in a rather straightforward way for
the downstream perturbations as a function of $\dXkp$, $\dVkp$ and the
upstream perturbations. In Sec.~\ref{sec:dscatt} and \ref{sec:utrans},
we solve a slightly different problem, by decomposing the downstream
perturbations over the Riemann invariants of the linearized MHD
equations; as shown in these Sections, one can then solve the above
equations for the outgoing wave modes and $\dXkp$, assuming harmonic
time dependence of $\dXkp$. In the following Sec.~\ref{sec:nonlin}, we
point out the existence of a particular non-perturbative solution to
the shock crossing equations, which is valid to all orders of
perturbations, in a 2D configuration ($k_z\,\rightarrow\,0$). Such a
solution may be particularly useful to set the initial data of a
numerical simulation of the evolution of the downstream at a
non-linearly corrugated shock wave.

\subsection{Non-linear corrugation in the 2D limit $k_z=0$}\label{sec:nonlin}
The previous section dealt with corrugation at first order in the
perturbations, thus assuming a linear regime, in which
$\vert\delta\ell\vert\,\ll\,1$, or equivalently
$\gfo^2\vert\dVkp/c\vert\,\ll\,1$ and $\gfo\vert
k_\perp\dXkp\vert\,\ll\,1$. One can actually obtain a solution at the
non-perturbative level in the particular case where the shock remains
smooth along the background magnetic field (which is assumed to be
aligned along the $\boldsymbol{z}$ direction). Since the present
analysis does not make any perturbative expansion or any Fourier
decomposition, it remains valid if the upstream quantities contain
spatial modulations transverse to the background magnetic field.

In contrast with the analyses of subsequent Sections, the present
analysis solves the shock crossing equations for the various fluid
quantities as a function of the shock normal, whose time and spatial
evolution dictate the amplitude of shock corrugation; however, it does
not specify how the latter is controlled by the past history of all
perturbations advected through the shock.

One then writes the flow four-velocity downstream
$u_2^\mu\,=\,\left(\gamma_2,\gamma_2\beta_{2,x},\gamma_2\beta_{2,y},\gamma_2\beta_{2,z}\right)$
and makes no particular assumption as to the magnitude of
$\gamma_2$. The magnetic field in the downstream plasma is written
$\boldsymbol{B}\,=\,\left(B_{2,x},B_{2,y},B_{2,z}\right)$.  Upstream
quantities remain unchanged. The crucial quantity is the shock normal,
which we write, in all generality, in the form:
\begin{equation}
\ell_\mu\,=\,\left(-\gamma_\ell \beta_{\ell,t},\gamma_{\ell},-\gamma_{\ell}\beta_{\ell,y},0\right)
\end{equation}
Of course, to preserve the space-like nature of the shock normal, this
four-vector must satisfy:
\begin{equation}
  \left\vert\beta_{\ell,t}\right\vert\,<\,\sqrt{1+\beta_{\ell,y}^2}
\end{equation}
The linear regime can be recovered through the substitution
\begin{eqnarray}
\gamma_{\ell}&\,\rightarrow\,&\gfo\left(1+\gfo^2\bfo\dVkp\right)\\
\gamma_{\ell}\beta_{\ell,t}&\,\rightarrow\,&\gfo\left(\bfo+\gfo^2\dVkp\right)\\
\gamma_{\ell}\beta_{\ell,y}&\,\rightarrow\,& i k_y\gfo\dXkp
\end{eqnarray}

This shock normal allows to describe a shock surface arbitrarily
rippled in the $y-$direction, with an arbitrary time behavior. It is
assumed of course that the scales over which these deformations take
place remain much larger than the thickness of the shock, so that the
shock crossing conditions can be applied at every point of the shock
surface.

\begin{widetext}
These shock crossing conditions then imply for the magnetic field components:
\begin{eqnarray}
B_{2,x}&\,=\,&0\\
B_{2,y}&\,=\,&0\\
B_{2,z}&\,=\,&B_1
\frac{\beta_1-\beta_{\ell,t}}{\beta_{2,x}-\beta_{\ell,t}-\beta_{2,y}\beta_{\ell,y}}\label{eq:nlB2z}
\end{eqnarray}
Regarding the velocity components, one finds a consistency condition
for $\beta_{2,z}$, while $\beta_{2,y}$ is written in terms of
$\beta_{2,x}$:
\begin{eqnarray}
\beta_{2,z}&\,=\,&0\\
\beta_{2,y}&\,=\,&\frac{\left(\beta_1-\beta_{2,x}\right)\beta_{\ell,y}}
{1-\beta_1\beta_{\ell,t}}
\end{eqnarray}
Finally, one can obtain equations for $W_2$ and $P_2$:
\begin{eqnarray}
W_2&\,=\,&W_1\gamma_1^2\frac{(\beta_1-\beta_{\ell,t})
(1-\beta_1\beta_{\ell,t})^2}{\gamma_2^2(1-\beta_{2,x}\beta_{\ell,t})\left[
(\beta_{2,x}-\beta_{\ell,t})(1-\beta_1\beta_{\ell,t})-(\beta_1-\beta_{2,x})\beta_{\ell,y}^2\right]}\\
P_2&\,=\,&W_1\gamma_1^2\frac{(\beta_1-\beta_{2,x})(\beta_1-\beta_{\ell,t})}
{1-\beta_{2,x}\beta_{\ell,t}}
\end{eqnarray}
These two equations neglect terms of order $P_1$ compared to
$W_1\gamma_1^2$, which corresponds to the usual strong shock
assumption. They can be combined with an equation of state
$\hat\gamma_2\,=\,4/3$, with Eq.~(\ref{eq:nlB2z}) and the
ultra-relativistic limit $\beta_1\,\simeq\,-1$ to derive a single
equation for $\beta_{2,x}$, which can be solved analytically:
\begin{equation}
3\frac{1+\sigma_1}{1-\beta_{2,x}\beta_{\ell,t}}+
\frac{\left(1+\sigma_1\right)\left(1+\beta_{\ell,t}\right)}
{\left(1+\beta_{\ell,t}\right)\left(\beta_{2,x}-\beta_{\ell,t}\right)+
\left(1+\beta_{2,x}\right)\beta_{\ell,y}^2}
-\sigma_1\left(1+\beta_{\ell,t}\right)\frac{\left(1+\beta_{\ell,t}\right)^2\left(1-\beta_{2,x}\right) -
  \left(1+\beta_{2,x}\right)\beta_{\ell,y}^2}
{\left[\left(1+\beta_{\ell,t}\right)\left(\beta_{2,x}-\beta_{\ell,t}\right)+
\left(1+\beta_{2,x}\right)\beta_{\ell,y}^2\right]^2}\,=\,0
\end{equation}

The root which matches the correct solution in the uncorrugated limit is:
\begin{equation}
\beta_{2,x}\,=\,\frac{A_{\beta_{2,x}}}{B_{\beta_{2,x}}}
\end{equation}
with
\begin{eqnarray}
A_{\beta_{2,x}}&\,=\,&
1 - 4 \beta_{\ell,t} - 10 \beta_{\ell,t}^2 - 4\beta_{\ell,t}^3 + \beta_{\ell,t}^4 + 
   7 \beta_{\ell,y}^2 - 7 \beta_{\ell,t}^2 \beta_{\ell,y}^2 + 6 \beta_{\ell,y}^4 + 
   2 \sigma_1 - 4 \beta_{\ell,t}^2 \sigma_1 + 2 \beta_{\ell,t}^4 \sigma_1 + 
   8 \beta_{\ell,y}^2 \sigma_1 -\nonumber\\
&&\quad 8 \beta_{\ell,t}^2 \beta_{\ell,y}^2 \sigma_1 + 
   6 \beta_{\ell,y}^4 \sigma_1 - (1 + \beta_{\ell,t})^2 (-1 + \beta_{\ell,t}^2 - \
\beta_{\ell,y}^2)\sqrt{1 + 16 \sigma_1 (1 + \sigma_1)}\nonumber\\
B_{\beta_{2,x}}&\,=\,&
2 \left\{\beta_{\ell,t}^4 \sigma_1 - 
     3 (1 + \beta_{\ell,y}^2)^2 (1 + \sigma_1) + \beta_{\ell,t}^3 (1 + 
        4 \sigma_1) - \beta_{\ell,t} (1 + \beta_{\ell,y}^2) (5 + 
        4 \sigma_1) + \beta_{\ell,t}^2 \left[-1 + \beta_{\ell,y}^2 + 
        2 (1 + \beta_{\ell,y}^2) \sigma_1\right]\right\}\nonumber\\
&&
\end{eqnarray}
Note indeed that in the limit $\beta_{\ell,t}\,\rightarrow\,0$,
$\beta_{\ell,y}\,\rightarrow\,0$ and $\gamma_{\ell}\,\rightarrow\,1$,
one recovers the unperturbed shock front in the shock front frame
with, accordingly, $\beta_{2,x}\,\rightarrow\,-\bfo$.

\end{widetext}

\section{Scattering of downstream magnetosonic modes}\label{sec:dscatt}
By definition, a fast magnetosonic shock, as that in which we are
interested, propagates relatively to the upstream plasma at a velocity
larger than the largest velocity of plasma fluctuations of the
upstream plasma $\beta_{\rm FM\vert 1}$. Relative to the downstream
plasma, however
\begin{equation}
\beta_{\rm A\vert2}\,<\,\bfo\,<\,\beta_{\rm FM\vert2}
\end{equation}
where $\beta_{\rm FM\vert2}\,=\,\left(\beta_{\rm A\vert2}^2+c_{\rm
  s\vert2}^2-\beta_{\rm A\vert2}^2c_{\rm s\vert2}^2\right)^{1/2}$
represents the group velocity of fast magnetosonic waves propagating
along the shock normal; $c_{\rm s}$ denotes the sound velocity and
$\beta_{\rm A}$ the (relativistic) Alfv\'en $3-$velocity, see
App.~\ref{sec:appWM}.  This implies that the far downstream plasma is
in causal contact with the shock front through the exchange of fast
magnetosonic waves only; conversely, all waves emitted by the shock
front impact the downstream, of course.

\begin{figure}[t]
\includegraphics[width=0.9\columnwidth]{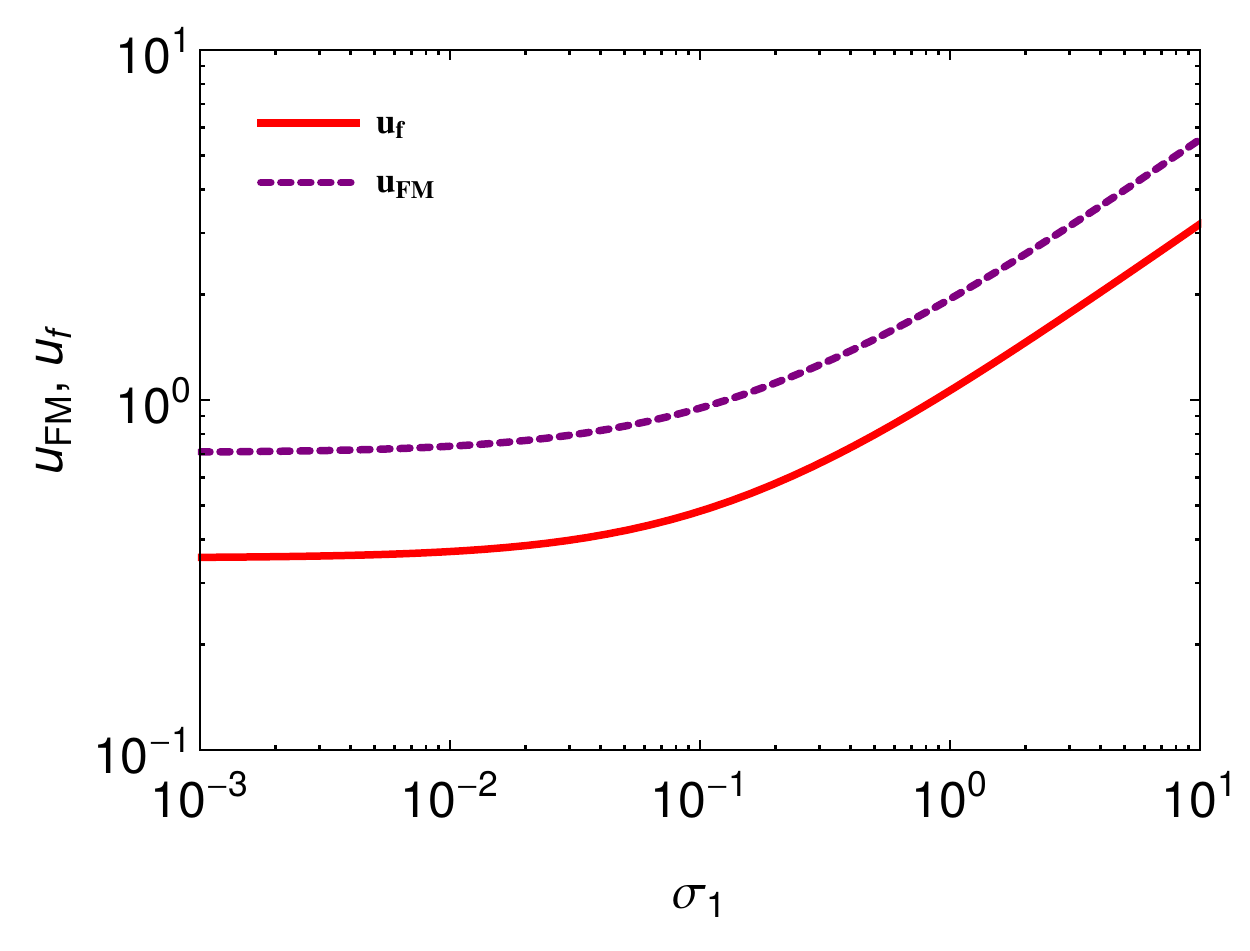}
\caption{The four-velocities of the shock front $u_{\rm
    f}\,\equiv\,\gfo\bfo$ and of the fastest magnetosonic mode $u_{\rm
    FM}\,\equiv\,\beta_{\rm FM}/\left(1-\beta_{\rm FM}^2\right)^{1/2}$
  of the downstream plasma as a function of $\sigma_1$, in the
  ultra-relativistic limit.
  \label{fig:betavel}}
\end{figure}

For completeness, we show in Fig.~\ref{fig:betavel} the
four-velocities of the shock front $u_{\rm f}\,\equiv\,\gfo\bfo$ and
of the fastest magnetosonic mode $u_{\rm FM}\,\equiv\,\beta_{\rm
  FM}/\left(1-\beta_{\rm FM}^2\right)^{1/2}$ as a function of the
magnetization of the upstream plasma, in the ultra-relativistic limit
$\gamma_1\,\gg\,1$.

\begin{figure}[t]
\includegraphics[width=0.9\columnwidth]{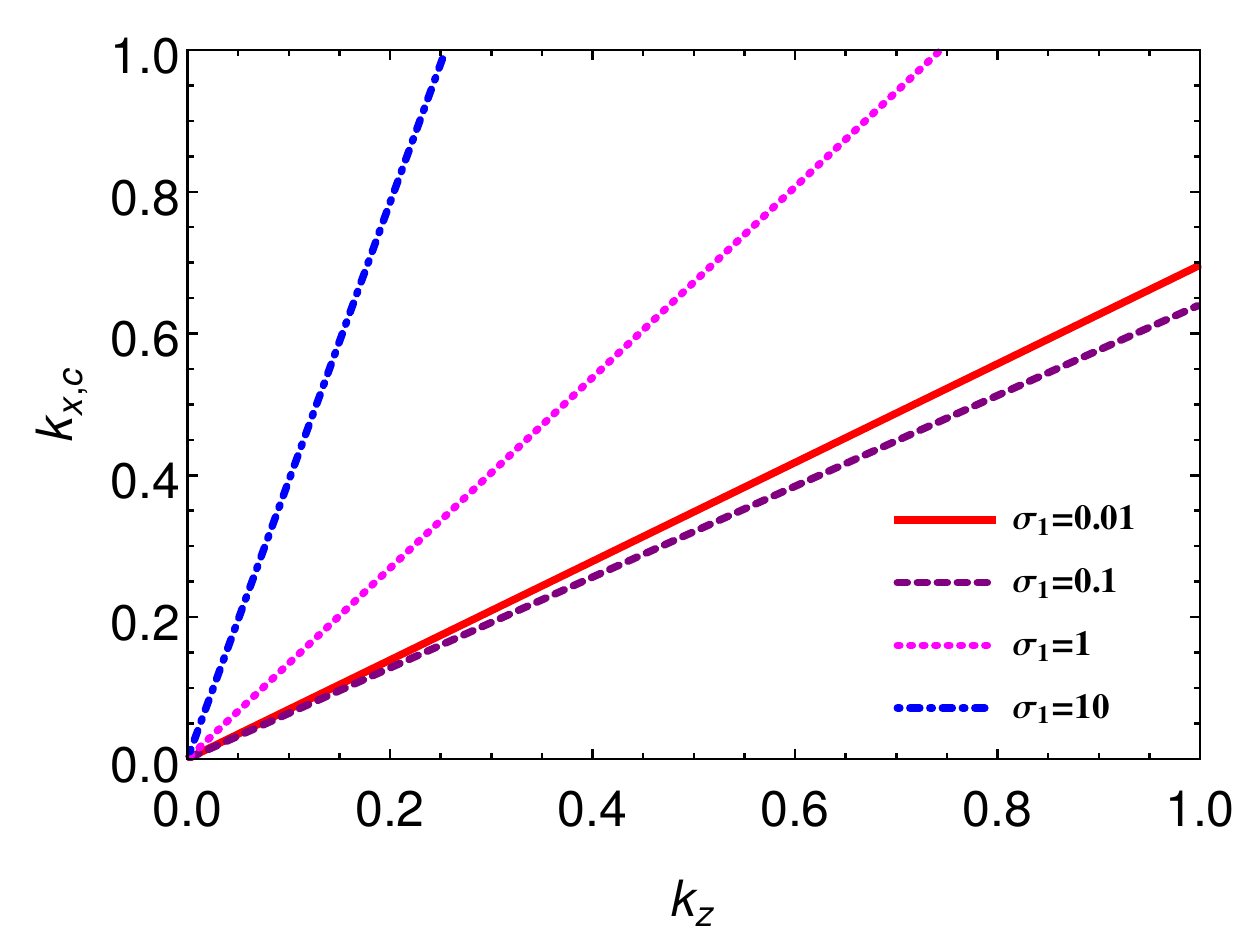}
\caption{The minimum value of $k_x$ as a function of $k_z$ such
  that the downstream fast magnetosonic mode outruns the shock, for
  various values of $\sigma_1$, in the ultra-relativistic limit
  $\gamma_1\,\gg\,1$, assuming $k_y\,=\,0$.  \label{fig:betacrit}}
\end{figure}

For a generic wave vector $\boldsymbol{k}$ (downstream frame), the
group velocity of downstream fast magnetosonic waves is
$\boldsymbol{\beta_{\rm g,FM\vert 2}}\,\equiv\,{\rm d}\omega_{\rm
  FM\vert2}/{\rm d}\bsk$. At given values of $(k_y,k_z)$, there thus
exists a critical value of $k_{x,\rm c}$ above which the $x-$
component $\beta_{\rm g,FM\vert2,x}\,>\,\bfo$. This value is
shown in Fig.~\ref{fig:betacrit} for various values of the
magnetization, as a function of $k_z$, assuming $k_y\,=\,0$; for
$k_y\,\neq\,0$, the minimum value of $k_x$ is typically raised by
$\,\sim\,k_y$ with respect to those shown in
Fig.~\ref{fig:betacrit}. As this figure shows, $k_{x,\rm c}$ becomes
large at large values of $\sigma_1$, because the shock velocity $\bfo$
becomes itself large; correspondingly, a smaller fraction of the phase
space of turbulence is in contact with the shock at larger values of
$\sigma_1$.

Provided $\beta_{\rm g,FM\vert2,x}\,>\,\bfo$, downstream fast
magnetosonic waves thus lead to the corrugation of the shock front. In
order to obtain an analytical description of the downstream turbulence
in the shock vicinity, one can then solve the problem as follows:
given one incoming fast magnetosonic wave outrunning the shock,
represented by a particular combination of the MHD perturbations of
the downstream, we determine the outgoing waves, namely one fast
magnetosonic wave, two Alfv\'en waves, two slow magnetosonic waves and
one entropy wave, as well as the shock corrugation $\dXkp$ (as
discussed below, an assumption of stationarity then fixes
$\dVkp\,=\,-i\,\omega_{\rm f}\dXkp$). In the present discussion, the
upstream is assumed unperturbed, so all terms $R_i$, $i\,=\,1\dots 9$,
vanish in Eqs.~(\ref{eq:sysbound}).

In order to solve the system, we first need to specify the wave
characteristics. In a stationary regime, the shock front reacts
harmonically to the excitation by an incoming fast magnetosonic wave
with a frequency (defined in the downstream rest frame):
\begin{equation}
\omega_{\rm f}\,=\,\omega_{<}-\beta_{\rm f}k_{x,<}
\end{equation}
in terms of $\omega_<$, the frequency of the incoming fast
magnetosonic wave and $k_{x,<}$ its $x$ wavenumber. Consequently,
$\dVkp\,=\,-i\,\omega_{\rm f}\dXkp$. This relation is a direct
expression of the Doppler effect associated to the motion of the shock
front relatively to downstream: the incoming wave indeed behaves as
$\delta\psi_{k,<}\,\sim\,\exp\left[- i \omega_{<}t+ik_{x,<}x +
  i\bskp\cdot\boldsymbol{x_\perp}\right]$, so that on the shock front
where $x\,=x_{\rm f}\,=\,\bfo t$, $\delta\psi_{k,<}\left[x_{\rm
    f}(t),t\right]\,\propto\,\exp\left[-i\omega_{\rm f}t+
  i\bskp\cdot\boldsymbol{x_\perp}\right]$.

Correspondingly, outgoing waves obey the relation
\begin{equation}
\omega_{i}-\beta_{\rm f}k_{x,i}\,=\,\omega_{\rm f}\label{eq:omegamatch}
\end{equation}
where $i$ denotes the wave mode. Since $\omega_{i}$ depends on
$k_{x,i}$, while $k_y$ and $k_z$ remain unchanged in the scattering
process, the above equation determines $k_{x,i}$, hence $\omega_{i}$
for each mode. The various plane wave modes thus all oscillate at the
same frequency on the shock surface and share the same wavenumber in
the transverse directions; however, due to their differing dispersion
relations, they exhibit different frequencies and $x-$ wavenumbers in
the downstream plasma. Once $k_{x,<}$, $k_y$ and $k_z$ have been
specified, all frequencies and $x-$ wavenumbers are determined
uniquely.

Formally, the problem can be written in terms of a linear response
relating the amplitude of the incoming wave to that of the outgoing
waves and of the shock corrugation. Concerning the latter, one can
write
\begin{eqnarray}
\delta X&\,=\,&\int\frac{{\rm d}^2k_\perp{\rm d}\omega_{\rm
  f}}{(2\pi)^3} \delta X_{\omega_{\rm f},\bskp}e^{-i\omega_{\rm
      f}t+i\bskp\cdot\boldsymbol{x_\perp}}\nonumber\\
&\,=\,& \int\frac{{\rm
      d}^2k_\perp{\rm d}k_{x,<}}{(2\pi)^3}e^{-i\omega_{\rm
        f}t+i\bskp\cdot\boldsymbol{x_\perp}} {\cal T}_{X,\bsk}\delta\psi_{<,\bsk}
\end{eqnarray}
introducing the transfer function
\begin{equation}
{\cal T}_{X,\bsk}\,\equiv\,\left\vert\frac{{\rm d}\omega_{\rm f}}{{\rm
    d}k_{x,<}}\right\vert\,\frac{\delta X_{\omega_{\rm f},\bskp}}{\delta\psi_{<,\bsk}}
\end{equation}

Regarding the variables describing the perturbations of the downstream
plasma, one must first decompose them as a sum over the wave modes,
i.e. over the eigenmodes of the system of linearized MHD
equations. This is done through the matrix $\boldsymbol{{\cal M}}$,
whose columns are the eigenvectors of the linearized MHD equations, as
described in App.~\ref{sec:appWM}. Recalling the notation introduced
in that appendix, $\boldsymbol{\delta\xi}$ represents the set of 8
perturbation variables $\left(\delta n,\delta
p,\boldsymbol{\delta\beta},\boldsymbol{\delta B}\right)$, while
$\boldsymbol{\delta\psi}$ represents the set of 8 wave modes of
linearized MHD; one of these 8 modes is an unphysical ghost mode
carrying non-vanishing $\boldsymbol{\nabla}\cdot\boldsymbol{B}$, which
must be included for a formal closure of the system, but which is not
excited by the interaction of turbulence with the shock front, once
the shock crossing conditions have been properly written.

For $\delta\xi_i\,\in\,\left\{\delta n_2,\, \delta p_2,\,
\boldsymbol{\delta\beta}_2,\,\boldsymbol{\delta B}_2\right\}$
($i=1,\ldots,8$), the decomposition introduced in App.~\ref{sec:appWM}
takes the form
\begin{equation}
  \delta \xi_i\,=\,\sum_j\,\int\frac{{\rm d}k_{x,j}{\rm d}k_y{\rm d}k_z}{(2\pi)^3}
  \,\,e^{-i\omega_jt+ik_{x,j}x+i\bskp\cdot\boldsymbol{x_\perp}}
\boldsymbol{\mathcal M}_{ij}\,\delta\psi_{j,\bsk}
\end{equation}
the sum over $j$ running over the 8 modes.  It is understood here that
the wave vectors and frequencies satisfy the matching conditions
discussed above. Furthermore, one of the 8 modes is actually the
incoming fast magnetosonic mode $\delta\psi_{<,\bsk}$, with $x-$
wavenumber $k_{x,<}$ and frequency
$\omega_{j}\,=\,\omega_{<}$. Defining the transfer functions, with the
index $j$ ranging over the wave modes
\begin{equation}
{\cal T}_{j,\bsk}\,\equiv\,\left\vert\frac{{\rm d}k_{j,x}}{{\rm
    d}k_{x,<}}\right\vert\,\frac{\delta \psi_{j,\bsk}}{\delta\psi_{<,\bsk}}
\end{equation}
with ${\cal T}_{<,\bsk}\,=\,1$ by definition, one can rewrite the
above as
\begin{eqnarray}
\delta \xi_i&\,=\,&\int\frac{{\rm d}^2k_\perp{\rm d}k_{x,<}}{(2\pi)^3}\,
\sum_{j}e^{-i\omega_jt+i\bskp\cdot\boldsymbol{x_\perp}+ik_{x,j}x}\nonumber\\
&&\quad\boldsymbol{\mathcal M}_{ij}\,{\cal T}_{j,\bsk}\delta\psi_{<,\bsk}
\end{eqnarray}
which provides a formal solution to the scattering problem once the
transfer functions have been determined.

The system Eq.~(\ref{eq:sysbound}) can be written formally as
\begin{equation}
\boldsymbol{\mathcal S}_{\bskp}\cdot\,^{\rm T}\left\{\delta X_{\bskp},\,\boldsymbol{\delta \xi}_{\bskp}\right\}\,=\,0\label{eq:s2a}
\end{equation}
where $\boldsymbol{\mathcal S}_{\bskp}$ is an $8\times9$ matrix and where the
perturbations $\boldsymbol{\delta \xi}_{\bskp}$ represent the 2d
Fourier transform of $\boldsymbol{\delta\xi}$ in
$\boldsymbol{x_\perp}$, as evaluated on the unperturbed shock surface
$\overline\Phi(x,t)\,=\,0$. The matching conditions for the wave
frequencies and parallel wavenumbers guarantee that all the wave modes
and $\delta X_{\bskp}$ share the same time and transverse spatial
dependence on this shock surface.  Using the decomposition, on the
unperturbed shock surface
\begin{equation}
\delta \xi_{i,\bskp}\,=\,e^{-i\omega_{\rm f}t+i\bskp\cdot\boldsymbol{x_\perp}}\,\sum_j
\boldsymbol{\mathcal M}_{ij}\delta\psi_{j,\bsk}
\end{equation}
one can bring the system (\ref{eq:s2a}) into an equivalent form:
\begin{equation}
\boldsymbol{\mathcal R}_{\bsk}\cdot\,^{\rm T}\left\{\delta X_{\omega_{\rm f},\bskp},\,\boldsymbol{\delta \psi}_{>,\bsk}\right\}\,=\,
\boldsymbol{R_{<,\bsk}}\delta\psi_{<,\bsk}\label{eq:s2b}
\end{equation}
where the index $_{>}$ indicates that the sum runs over the outgoing
modes only. Since MHD has 8 wave modes (including one unphysical ghost
mode, see App.~\ref{sec:appWM}), and since the column
associated to the incoming mode has been extracted out of
$\boldsymbol{\mathcal R}_{\bsk}$ and sent into the r.h.s., the matrix
$\boldsymbol{\mathcal R}_{\bsk}$ is now $8\times 8$ and the above
linear problem allows to solve for $\delta X_{\bskp}$ and
$\boldsymbol{\delta\psi}_{>,\bskp}$ in terms of $\delta
\psi_{<,\bskp}$.

One can write an analytical solution of the above system, but given
the large rank of the matrix, its inverse cannot be written in a
compact way. For these reasons, we provide in the following direct
estimates of the solutions for various cases of interest.

\subsection{Results}
\begin{figure}[h!]
  \includegraphics[width=0.9\columnwidth]{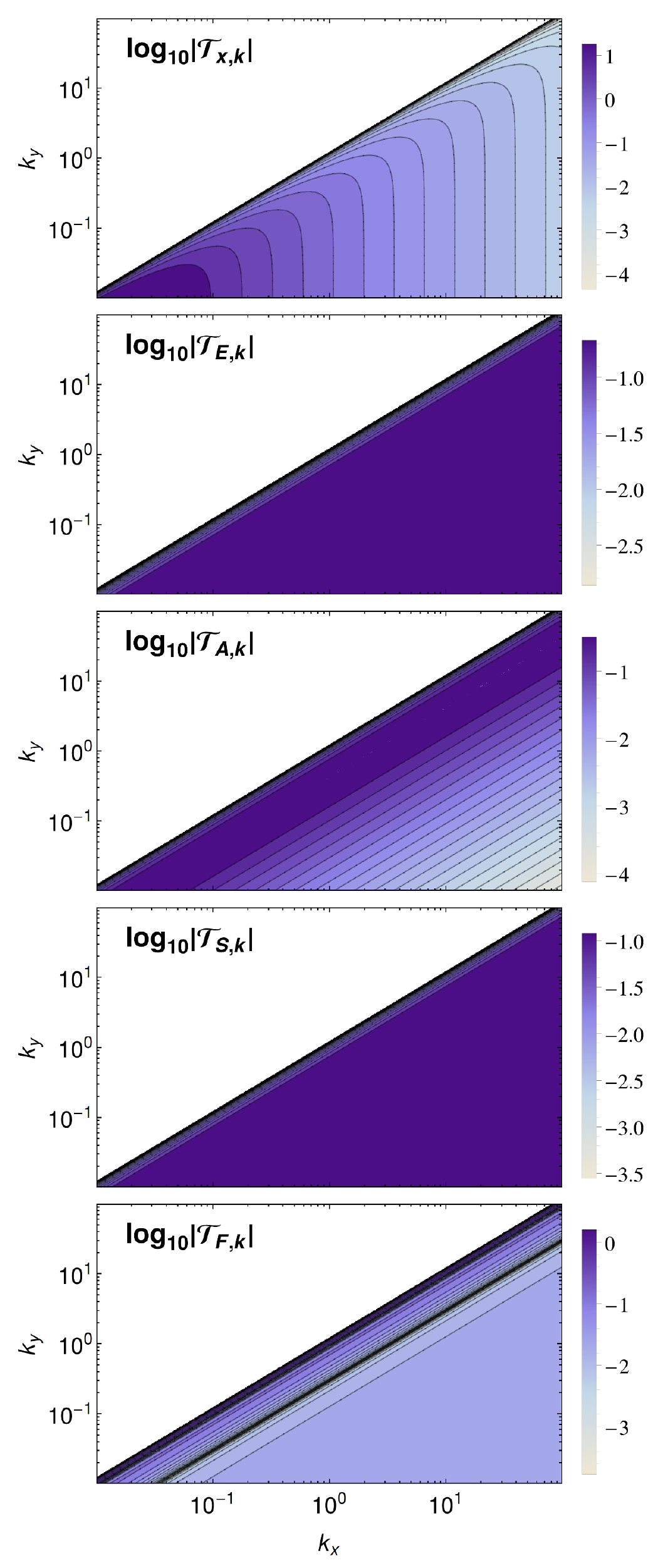}
\caption{Contour plots of the transfer functions for
  $\sigma_1\,=\,0.1$, $\gamma_1\,=\,10^4$ (ultra-relativistic limit)
  in the 2D-plane $(k_x,\,k_y)$; $k_z\,=\,10^{-3}$ everywhere
  here. The symbols are as follows: $\mathcal{T}_{X,\bsk}$ represents
  the response of the shock corrugation amplitude, and indices $_{\rm
    E}$, $_{\rm A}$, $_{\rm S}$ and $_{\rm F}$ respectively refer to
  the entropy mode, the two Alfv\'en modes (sum of the two responses),
  the two slow magnetosonic modes (also summed) and the reflected fast
  magnetosonic mode.
\label{fig:trans2dy}}
\end{figure}

In Fig.~\ref{fig:trans2dy}, we show a contour plot of the transfer
functions for the various modes, in the plane $(k_{x,<},k_y)$, for
$\sigma_1\,=\,0.1$ and $\gamma_1\,=\,10^4$ (representative of the
ultra-relativistic limit $\gamma_1\,\gg\,1$). In this figure,
$k_z\,=\,10^{-3}$; units for wavenumbers are arbitrary, as the problem
of corrugation of a planar shock front does not possess an intrinsic
scale. For the two slow magnetosonic and for the two Alfv\'en modes,
the transfer functions have been respectively added together.

\begin{figure}[h!]
  \includegraphics[width=0.9\columnwidth]{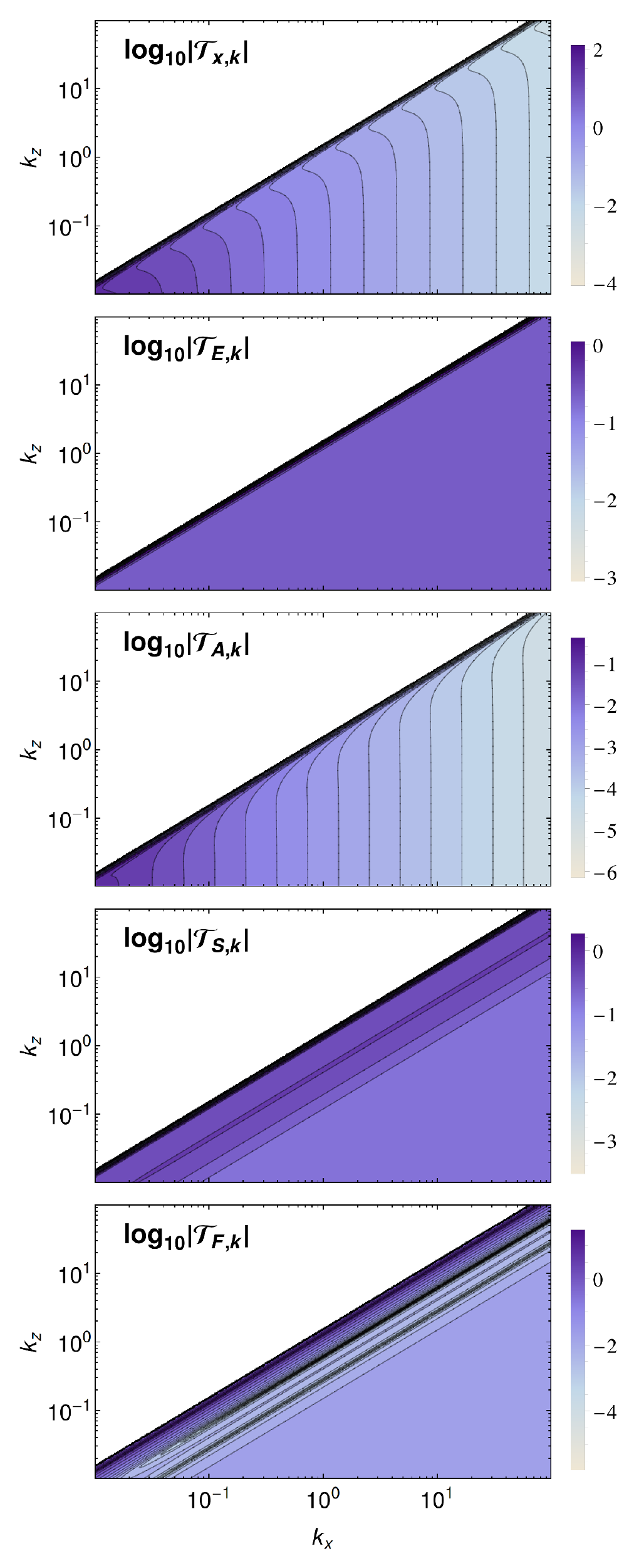}
\caption{Contour plots of the transfer functions for
  $\sigma_1\,=\,0.1$, $\gamma_1\,=\,10^4$ (ultra-relativistic limit)
  in the 2D-plane $(k_x,\,k_z)$; $k_y\,=\,10^{-3}$ everywhere here.
\label{fig:trans2dz}}
\end{figure}

Figure~\ref{fig:trans2dz} presents a contour plot equivalent to that
shown in Fig.~\ref{fig:trans2dy}, but for perturbations along the
magnetic field; i.e., $k_y\,=\,10^{-3}$ and the contour is shown in
the $(k_{x,<},k_z)$ plane.

A general trend observed in these figures and in a more systematic
survey is that the incoming fast magnetosonic mode is converted in
roughly similar proportions in the various outgoing modes.  At large
values of $k_x$, in particular $k_x\,\gg\,k_y,\,k_z$, the transfer
function for the shock corrugation amplitude scales as $1/k_x$, with
typically $\left\vert\mathcal{T}_{X,\bsk}\right\vert\,\sim\,{\mathcal
  O}(1)k_x^{-1}$. This scaling appears as a natural consequence of the
scale invariance of the problem at hand -- there being no natural
length scale associated to the physics of a planar infinite shock
front in the MHD limit -- once one recalls that
$\left\vert\mathcal{T}_{X,\bsk}\right\vert$ carries the dimensions of
a length scale, because $\delta X$ is a length scale while $\delta
\psi_<$ is dimensionless. The prefactor is typically of order unity,
although it depends somewhat on the nature of the incoming wave and on
the magnetization of the upstream plasma.

\section{Transmission of upstream turbulence}\label{sec:utrans}
This Section discusses the transmission of upstream turbulence through
the shock. For the sake of simplicity, this discussion is restricted
to the transmission of entropy and Alfv\'en waves, for which the
Riemann invariants of the linearized MHD system of a streaming plasma
can be written in a compact way. In principle, the problem can be
generalized directly to include the transmission of upstream
magnetonic waves. However, the analysis is here carried out in the
rest frame of the downstream plasma, with respect to which the
upstream plasma is drifting at relativistic speeds. In this case, the
Riemann invariants associated to magnetosonic wave modes take quite
complicated expressions, making the algebra cumbersome. The following
therefore focuses on entropy and Alfv\'en waves; a numerical example
of the impact of incoming fast magnetosonic waves will nevertheless be
provided in Fig.~\ref{fig:tupAslc}.

The procedure follows that of Sec.~\ref{sec:dscatt}. One first
decomposes the drifting upstream turbulence in its eigenmodes in
Fourier space. The entropy mode is characterized by $\delta
n_1\,=\,n_1\delta\psi_{\rm E1}$ and all other perturbations
$\boldsymbol{\delta\beta}_1\,=\,\boldsymbol{\delta B}_1\,=\,0$ (note
that $\delta p_1$ is no longer a perturbation variable since the
upstream plasma is considered cold). In the rest frame of the upstream
plasma, the eigenfrequency is $\omega_{\rm E1\vert u}\,=\,0$, so that
in the downstream frame: $\omega_{\rm E1}\,=\,\beta_1k_x$.

In terms of the perturbation amplitude $\delta\psi_{\rm A1}$, the
Alfv\'en modes are characterized by:
\begin{eqnarray}
  \delta n_1&\,=\,&\delta \beta_{1z}\,=\,0,\nonumber\\
  \delta \beta_{1x}&\,=\,& \frac{1}{\gamma_1^2}\delta\psi_{\rm A1}\,\nonumber\\
  \delta \beta_{1y}&\,=\,& \frac{\beta_1\omega_{\rm A1}-k_x}{k_y}\delta\psi_{\rm A1}\,\nonumber\\
  \delta B_{1x}&\,=\,& -\frac{k_z}{\gamma_1^2\left(\omega_{\rm A1}-\beta_1k_x\right)}\delta\psi_{\rm A1}\,\nonumber\\
  \delta B_{1y}&\,=\,& -\frac{k_z\left(\beta_1\omega_{\rm A1}-k_x\right)}{k_y\left(\omega_{\rm A1}-\beta_1k_x\right)}\delta\psi_{\rm A1}\,\nonumber\\
  \delta B_{1z}&\,=\,& \beta_1\delta\psi_{\rm A1}\label{eq:psiA1}
\end{eqnarray}
with frequency: $\omega_{\rm A1}\,=\,\beta_1k_x \pm \beta_{\rm
  A1}k_z/\gamma_1^2$. It is easy to verify that one recovers the
corresponding eigenmode for a plasma at rest, Eq.~(\ref{eq:eigenvec}),
in the limit $\beta_1\,\rightarrow\,0$, $\gamma_1\,\rightarrow\,1$.

The frequency of the corrugation amplitude $\delta X_{\bskp}$ is determined by the matching:
\begin{equation}
  \omega_{\rm f}\,=\,\omega_1-\bfo k_x
\end{equation}  
with $\omega_1\,=\,\omega_{\rm E1}$ or $\omega_{\rm A1}$ depending on
the source of the perturbations entering the shock. The corrugations
induced on the shock are then converted into downstream outgoing
perturbations. The frequency $\omega_i$ and wavenumbers $k_{x,i}$ of
these modes are determined as previously in terms of $\omega_{\rm f}$,
of course.

There are now six outgoing modes: 1 entropy, 2 Alfv\'en, 2 slow
magnetosonic and 1 fast magnetosonic mode. That only one fast
magnetosonic mode is excited is a non-trivial result by itself, which
deserves some discussion. At a given value of $\omega_{\rm f}$ --
equivalently, at a given value of the $x-$ component $k_x$ of the
incoming perturbation -- one can find the values of $k_{x,i}$ of the
magnetosonic waves which satisfy the frequency matching condition
Eq.~(\ref{eq:omegamatch}) by solving the equation:
\begin{eqnarray}
  \left(\omega_{\rm f}+\bfo k_{x,i}\right)^4
  &-&\left(\beta_{\rm FM}^2k^2+\beta_{\rm A}^2c_{\rm s}^2k_z^2\right)
  \left(\omega_{\rm f}+\bfo k_{x,i}\right)^2  + \nonumber\\
 &&\quad\beta_{\rm A}^2c_{\rm s}^2k^2k_z^2\,=\,0
\end{eqnarray}
which is nothing else but the dispersion relation of magnetosonic
waves, with the frequency replaced by $\omega_{\rm f}+\bfo k_{x,i}$;
it is understood here that $k^2=k_{x,i}^2+k_y^2+k_z^2$. This quartic
equation has at most four real solutions, then corresponding to two
fast and two slow magnetosonic waves. One finds that there exists a
critical value of the incoming $k_x$ (for $k_x\,>\,0$), below which
the above equation has two real solutions and a pair of complex
conjugate solutions, and above which the equation has four real
solutions. At the critical value of $k_x$, written $k_{x,\rm c}$ in
the following, the group velocity of the downstream fast magnetosonic
wave is very close or equal to the shock velocity, $\beta_{{\rm
    g,FM},x}\,\simeq\,\bfo$. For $k_x\,>\,k_{x,\rm c}$, one of the
fast magnetosonic waves has a group velocity in excess of $\bfo$,
while the other has a group velocity smaller than $\bfo$. Therefore,
in this case, only one fast magnetosonic wave (the slower) can be
excited by the corrugation. For $k_x\,<\,k_{x,\rm c}$, one of the
complex solutions has a positive imaginary part, which corresponds to
an unphysical solution with unbounded amplitude towards far downstream
($x\,\rightarrow\,-\infty$). In this case, we thus set this wave to
zero, and retain only the wave with negative imaginary part of
$k_{x,i}$, which physically describes a mode localized on the shock
front.

 Formally, the problem is then written as in the previous
Section, see Eq.~(\ref{eq:s2b}), except that the source of corrugation
is no longer the downstream fast magnetosonic mode, but rather the
incoming upstream perturbation, as indicated by the r.h.s. of
Eq.~(\ref{eq:sysbound}):
\begin{equation}
\boldsymbol{\mathcal R}_{\bsk}\cdot\,^{\rm T}\left\{\delta
X_{\omega_{\rm f},\bskp},\,\boldsymbol{\delta
  \psi}_{>,\bsk}\right\}\,=\,
\boldsymbol{R}_{1,\bsk}\delta\psi_{1,\bsk}\label{eq:s2c}
\end{equation}
with $\boldsymbol{R}_{1,\bsk}$ determined by Eq.~(\ref{eq:sysrhs}) and
the decomposition of the perturbations $\delta n_{1,\bskp}$ etc. in terms of
$\delta\psi_{1,\bsk}$, for each of the two cases studied here,
$\delta\psi_{1,\bsk}\,=\,\delta\psi_{{\rm E1},\bsk}$ or $\delta\psi_{{\rm A1},\bsk}$.

One can then define the transfer functions
\begin{equation}
{\mathcal T}_{X,\bsk}\,\equiv\,\left\vert\frac{{\rm d}\omega_{\rm f}}{{\rm
    d}k_{x,<}}\right\vert\,\frac{\delta X_{\omega_{\rm f},\bskp}}{\delta\psi_{1,\bsk}}\label{eq:txu}
\end{equation}
and
\begin{equation}
{\cal T}_{j,\bsk}\,\equiv\,\left\vert\frac{{\rm d}k_{j,x}}{{\rm
    d}k_{x,<}}\right\vert\,\frac{\delta \psi_{j,\bsk}}{\delta\psi_{1,\bsk}}\label{eq:tju}
\end{equation}
as before, now expressed relatively to the incoming upstream perturbation.

If perturbations are sourced by upstream fluctuations, one finds that
there exist values of $(k_x,k_y,k_z)$ for which ${\rm
  Det}\,\boldsymbol{\mathcal R}_{\bsk}\,=\,0$, which correspond to a
resonant response of the shock corrugation to the incoming
perturbation, with formally infinite corrugation amplitude. This
result stands in contrast with the case studied in the previous
Section, for which one could not find values of $(k_x,k_y,k_z)$ which
lead to a vanishing determinant; the difference lies of course in the
relationship which ties $\omega_{\rm f}$ to $k_{x,<}$, and which
differs between those two cases.

\subsection{Non-resonant response}
\begin{figure}[h!]
  \includegraphics[width=0.9\columnwidth]{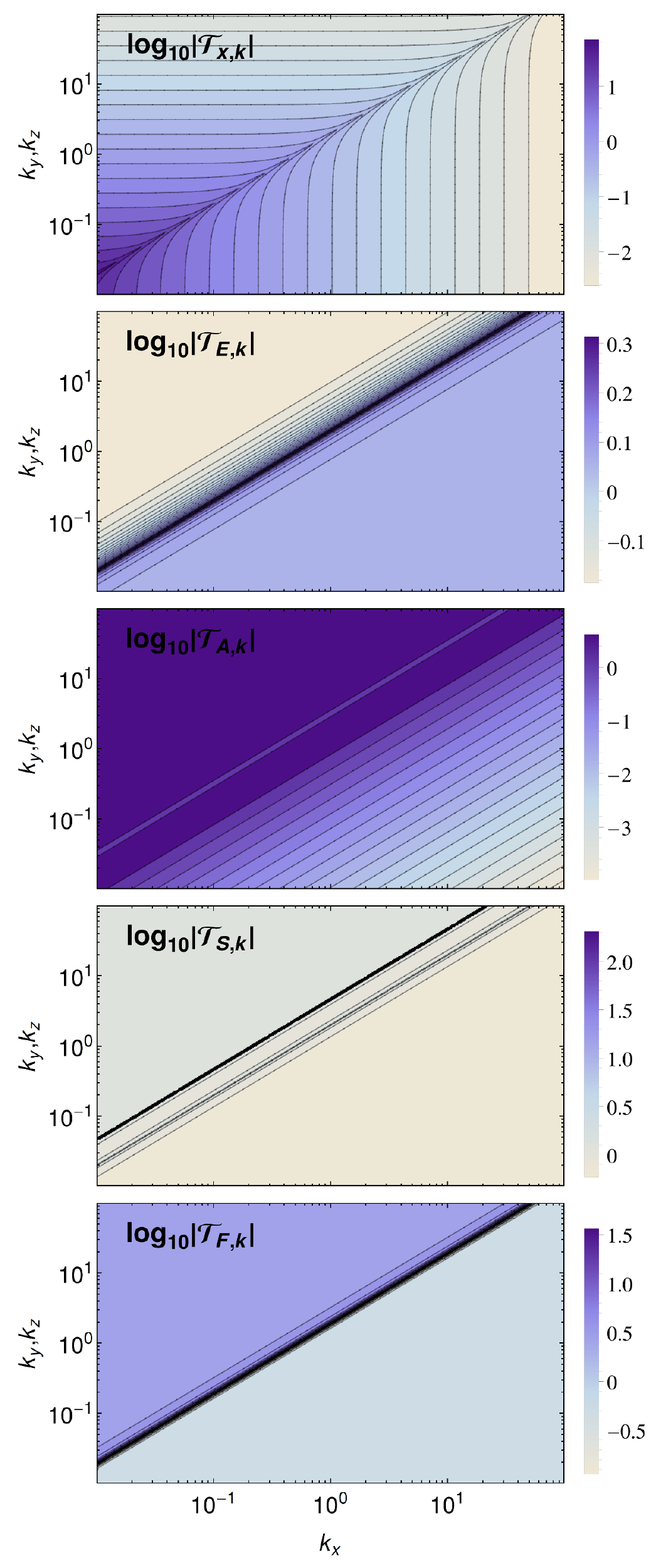}
\caption{Contour plots of the transfer functions for
  $\sigma_1\,=\,0.1$, $\gamma_1\,=\,10^4$ (ultra-relativistic limit)
  in the 2D-plane $(k_x,\,k_y)$ with $k_z\,=\,k_y$ everywhere
  here, for an incoming entropy mode
\label{fig:tupE}}
\end{figure}

\begin{figure}[h!]
  \includegraphics[width=0.9\columnwidth]{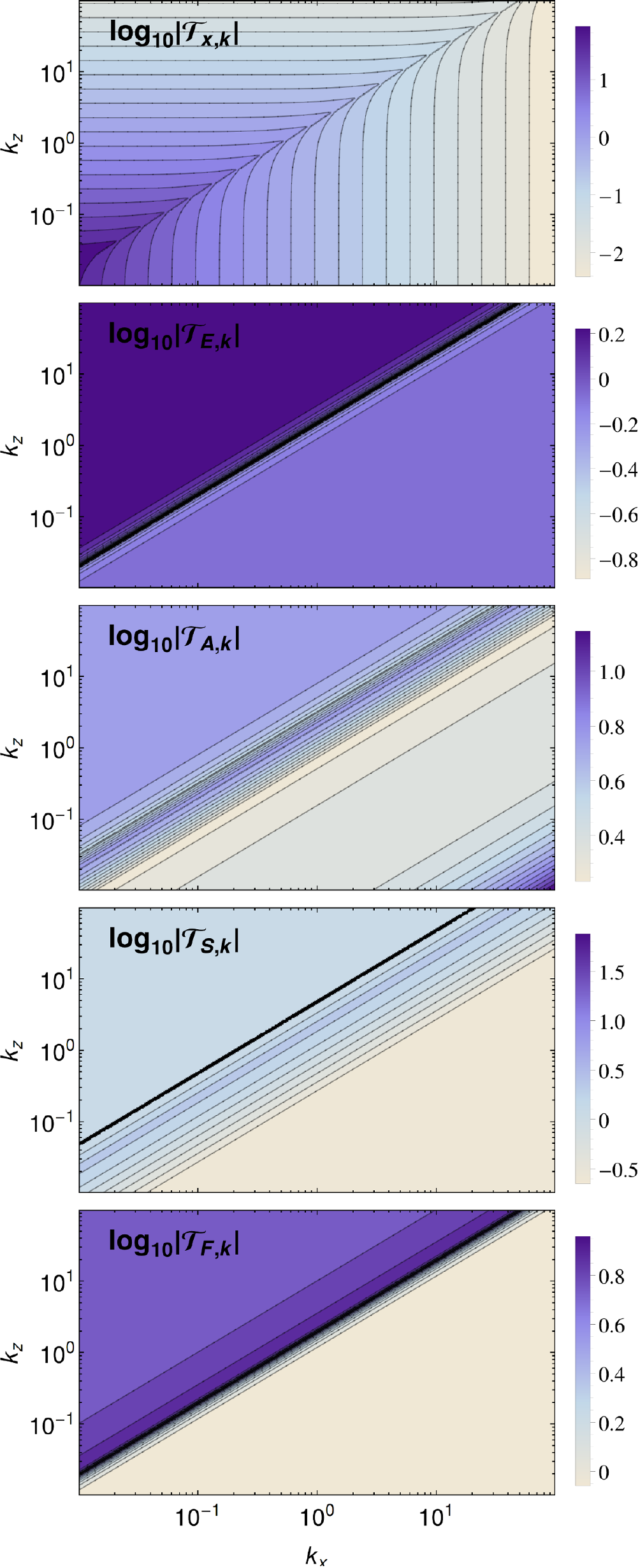}
  \caption{Contour plots of the transfer functions for
    $\sigma_1\,=\,0.1$, $\gamma_1\,=\,10^4$ (ultra-relativistic limit)
    in the 2D-plane $(k_x,\,k_y)$ with $k_z\,=\,k_y$ everywhere here,
    for an incoming Alfv\'en wave.
\label{fig:tupA}}
\end{figure}

As in Sec.~\ref{sec:dscatt}, we show here the transfer functions for
the response to the excitation by incoming entropy and Alfv\'en
modes. Given the large dimensionality of the parameter space, we
restrict these plots to the region in which $k_y\,=\,k_z$, to
$\sigma_1\,=\,0.1$ and to the ultra-relativistic limit
$\gamma_1\,\gg\,1$ (for practical matters, $\gamma_1\,=\,10^4$
here). The transmission of entropy modes is shown in
Fig.~\ref{fig:tupE}, while the transmission of Alfv\'en modes is shown
in Fig.~\ref{fig:tupA}.

These curves reveal a ridge along which the amplification takes large
values; this resonant response is analyzed in greater detail in the
following Sec.~\ref{sec:upres}. In other parts of parameter space, the
response of the shock front to the incoming perturbation is of order
unity, meaning $\left\vert k\delta X_{\bsk}\right\vert\,\sim\,{\cal
  O}(1)\,\delta\psi_1$, leading to the non-linear regime of
corrugation if the incoming amplitude is of order unity.

\subsection{Resonant response}\label{sec:upres}
As discussed above, at certain locations of $\bsk$, the determinant of
the response matrix of the shock corrugation and outgoing wave
amplitudes takes small or even vanishing values, leading to a large
amplification of the incoming perturbation in the present linear
approximation. For all cases surveyed, it was found that, for a given
pair $(k_y,k_z)$, there is at most one resonant value, written
$k_{x,\rm r}$ with $k_{x,\rm r}\,\simeq\,k_{x,\rm c}$ to very good
accuracy. The latter remark motivates the following interpretation: as
$k_x\,\rightarrow\,k_{x,\rm c}$, the outgoing fast magnetosonic wave
rides along with the shock front, because its group velocity matches
$\bfo$; therefore, the large corrugation and consequent amplification
of downstream modes follow from the build-up of the fast magnetosonic
mode energy on the shock.

\begin{figure}
  \includegraphics[width=0.9\columnwidth]{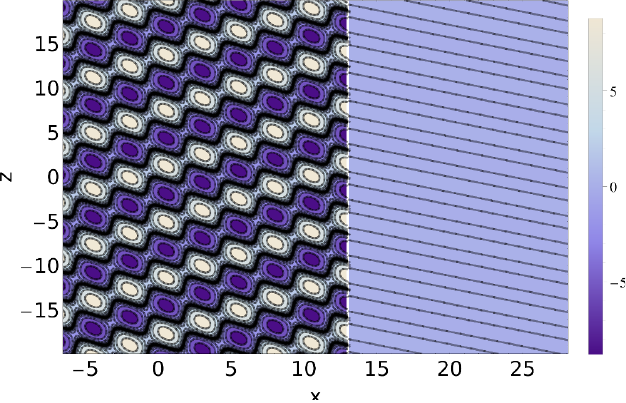}
\caption{Snapshot contour plot of the perturbation $\delta B_z$ in
  configuration space, in the $(x,y)$ plane. The shock front lies at
  $x\,=\,13$. An Alfv\'en perturbation with amplitude $\delta\psi_{{\rm
    A1},\bsk}\,=\,0.1$ enters the shock from upstream ($x\,>\,13$) and
  seeds corrugation; the amplitude of the outgoing mode reaches here
  $\vert\delta B_{z,\bskp}\vert\,\sim\,9$. Parameters are:
  $\sigma_1\,=\,0.1$, $\gamma_1\,=\,10^4$, $k_x\,=\,0.33$,
  $k_y\,=\,0.001$, $k_z\,=\,1$.
\label{fig:tupmap}}
\end{figure}

On a root of ${\rm Det}\,\boldsymbol{\mathcal R}_{\bsk}$, all
downstream perturbations diverge. In order to illustrate the effect of
such a resonant response, we plot in Fig.~\ref{fig:tupA} the $\delta
B_z$ perturbation, assuming an incoming perturbation $\delta\psi_{{\rm
  A1},\bsk}\,=\,0.1$ (hence $\vert\delta
B_{z,1,\bskp}\vert\,\simeq\,0.1$). Here, $k_x\,\simeq\,0.3308$
($-0.01\,$\% away from the actual root), $k_y\,=\,0.001$, $k_z\,=\,1.$,
$\sigma_1\,=\,0.1$ and $\gamma_1\,=\,10^4$. The contours are spaced
logarithmically; note the amplification by a factor $\gtrsim\,80$ of
the downstream perturbation. For these values, one finds a corrugation
amplitude $\left\vert\delta X_{\bskp}\right\vert\,\sim\,150$, well into
the non-linear regime; Fig.~\ref{fig:tupmap} should thus be considered
as an illustration, in the framework of the linear approximation.

\begin{figure}
  \includegraphics[width=0.9\columnwidth]{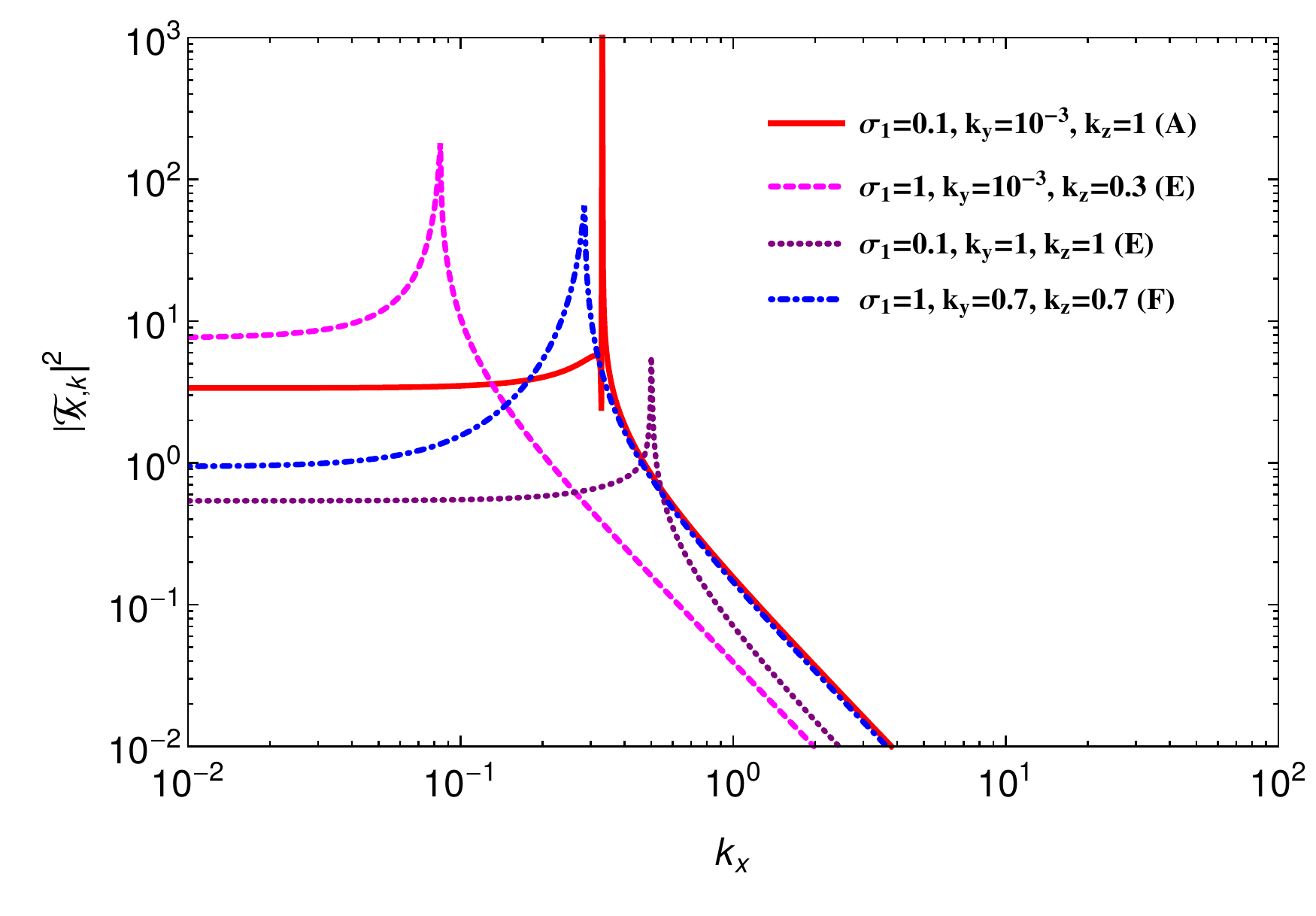}
\caption{Slice of the transfer function for various waves as
  indicated, $\gamma_1\,=\,10^4$ (ultra-relativistic limit) as a
  function of $k_x$; (A) refers to Alfv\'en waves, (E) to entropy
  waves and (F) to fast magnetosonic waves.
\label{fig:tupAslc}}
\end{figure}

Our study of parameter space reveals that the response at
$k_x\,\rightarrow\,k_{x,\rm r}$ dominates over that at other
wavenumbers. In order to illustrate this point, we show in
Fig.~\ref{fig:tupAslc} the modulus squared of the transfer function
for the corrugation amplitude as a function of $k_x$, for several
cases. This quantity relates the power spectrum of corrugation to the
power spectrum of the incoming turbulence, through
\begin{equation}
  {\mathcal P}_{\delta X,\bsk}\,=\,\left\vert{\mathcal T}_{X,\bsk}\right\vert^2 {\mathcal P}_{1,\bsk}
\end{equation}
with
\begin{equation}
  \left\langle\delta X^2\right\rangle\,=\,\int\frac{{\rm d}^3k}{(2\pi)^3}{\mathcal P}_{\delta X,\bsk}
\end{equation}
and similarly for the incoming wave in terms of
$\left\langle\delta\psi_1^2\right\rangle$. Figure~\ref{fig:tupAslc}
assumes an Alfv\'en wave in the incoming state.
The resonant behavior at $k_{x,\rm r}$ is clear in the above
figure. At larger values of $k_x$, in particular
$k_x\,\gg\,k_y,\,k_z$, one recovers the scaling
$\left\vert\mathcal{T}_{X,\bsk}\right\vert\,\sim\,{\mathcal
  O}(1)k_x^{-1}$ observed in Sec.~\ref{sec:dscatt}.

Figure~\ref{fig:tupAslc} also reveals that the magnitude of the
amplification at the resonant value of $k_x$ depends on the
parameters, in particular $k_y,\,k_z$ and $\sigma_1$. One typically
observes that for $k_y\,\gtrsim\,k_z$ or $\sigma_1\,\gtrsim\,0.3$, the
determinant does not vanish at $k_x\,\rightarrow\,k_{x,\rm r}$ but
takes a minimum value, nevertheless leading to large but finite
amplification of the incoming wave, while at $k_y\,\ll\,k_z$ and
$\sigma_1\,\lesssim\,0.3$, actual roots of ${\rm
  Det}\,\boldsymbol{\mathcal R}_{\bsk}\,=\,0$ exist, with
correspondingly infinite amplification.

For the latter case, a careful study of the transfer function in the
vicinity of the root, $k_x\,=\,k_{x,\rm r}(1+\epsilon)$, reveals that
\begin{equation}
  \left\vert\mathcal{T}_{X,\bsk}\right\vert\,\propto\,\frac{1}{\left\vert\epsilon\right\vert}
\end{equation}
which indicates that the power spectrum of corrugation diverges as
$1/\epsilon^2$ at the resonance. The present linear theory thus
strongly suggests that these resonances should dominate the response
of the shock to the incoming turbulence. One should nevertheless
recall that this analysis assumes a stationary upstream turbulence,
and that in the presence of modes with finite damping coefficients,
the influence of the above narrow resonances might be diminished. We
defer to a future work a detailed study of these resonances and their
consequences for the phenomenology of astrophysical shock waves.

\section{Conclusions}
This paper has provided a general analysis of the corrugation of
relativistic magnetized (fast) shock waves induced by the interaction
of the shock front with moving disturbances. Two cases have been
analyzed, depending on the nature of these disturbances: whether they
are fast magnetosonic waves originating from the downstream side of
the shock front, outrunning the shock, or whether they are eigenmodes
of the upstream plasma.

Working to first order in the perturbations of the flows, on both
sides of the shock front, as in the amplitude of corrugation $\delta
X$ of the shock front, we have provided transfer functions relating the
amplitude of the outgoing wave modes to the amplitude of the incoming
wave, thus developing a linear response theory for the corrugation. We
have then provided estimates of these transfer functions for different
cases of interest.

One noteworthy result is that the front generically responds with
$\vert\partial\delta X\vert \,\sim\,{\cal O}(1)\delta\psi_{<}$, where
$\delta\psi_{<}$ represents the amplitude of the incoming wave, the
partial derivative being taken along $t$, $y$ or $z$. The corrugation
remains linear as long as $\vert\partial\delta X\vert\,\ll\,1$,
therefore the present analysis is restricted to small amplitude
incoming waves. Interestingly, however, the extrapolation of the
present results indicates that non-linear corrugation can be achieved
in realistic situations.

In this respect, we have obtained an original solution for the
equations of shock crossing, in the non-perturbative regime; this
solution allows to calculate the amplitude of fluctuations in density,
pressure, velocity and magnetic field components at all locations (and
all times) on a shock surface which is arbitrarily rippled in the $y-$
direction, but smooth along the background magnetic field.

Furthermore, when corrugation is induced by upstream wave modes, we
find that there exist resonant wavenumbers $\bsk$ where the linear
response becomes large or even formally infinite, leading to large or
formally infinite amplification of the incoming wave amplitude and of
the shock corrugation. For a given pair $(k_y,k_z)$, there exists one
such resonance in $k_x$ for the incoming mode, corresponding to the
value at which the outgoing fast magnetosonic mode moves as fast as
the shock front. The physics of shock waves interacting with
turbulence containing such resonant wavenumbers should be examined
with dedicated numerical simulations able to probe the deep non-linear
regime, as the structure of the turbulence produced by the corrugation
may have profound consequences for our understanding of astrophysical
magnetized relativistic shock waves and their phenomenology.

\acknowledgments
  We thank A. Bykov, R. Keppens and G. Pelletier for insightful
  discussions and advice. This work has been financially supported by
  the Programme National Hautes \'Energies (PNHE) of the C.N.R.S. and
  by the ANR-14-CE33-0019 MACH project.





\appendix
\section{Wave modes in relativistic MHD}\label{sec:appWM}
The dynamics of the plasma is governed by the MHD equations:
\begin{eqnarray}
\partial_\mu\left(n u^\mu\right)&\,=\,&0\\
\partial_\mu T^{\mu\nu}&\,=\,&0\\
\partial_\mu\,^{\star}F^{\mu\nu}&\,=\,&0\\
\partial_\mu\,F^{\mu\nu}&\,=\,&4\pi j^\nu
\end{eqnarray}
The various quantities entering these equations are defined in
Sec.~\ref{sec:appS}.

The wave modes of an unbounded plasma can be obtained as usual by
linearizing these equations around an unperturbed state characterized
by uniform energy density, pressure and density, and regular magnetic
field, assumed oriented along $\boldsymbol{z}$:
$\boldsymbol{\overline{B}}=B_z \boldsymbol{z}$. The sound and
(relativistic) Alfv\'en velocities are defined according to:
\begin{equation}
c_{\rm s}^2\,=\,\hat\gamma\frac{p}{w},\quad
\beta_{\rm A}^2\,=\,\frac{\overline B^2}{4\pi W}
\end{equation}
and $\hat\gamma$ is the polytropic index.

As is well-known, the linearized equations and their solutions are
characterized, in Fourier space, by 8 perturbation variables
$\boldsymbol{\delta\xi}_{\bsk}\,=\,\left(\delta n_{\bsk},\delta
p_{\bsk},\boldsymbol{\delta\beta}_{\bsk},\boldsymbol{\delta
  B}_{\bsk}\right)$. These perturbations define 8 modes: one entropy
mode (indexed $_{\rm E}$), two Alfv\'en modes (indexed $_{\rm A}$),
two slow magnetosonic modes (indexed $_{\rm SM}$), two fast
magnetosonic modes (indexed $_{\rm FM}$), and one non-physical ghost
mode carrying non-vanishing $\boldsymbol{\nabla}\cdot\boldsymbol{B}$
(not discussed in the following). The corresponding frequencies of
these various modes are:
\begin{eqnarray}
\omega_{\rm E}&\,=\,&0,\nonumber\\
\omega_{\rm A}&\,=\,&\pm \beta_{\rm A}k_z,\nonumber\\
\omega_{\rm SM}&\,=\,& \pm\frac{1}{\sqrt{2}}\biggl\{\beta_{\rm FM}^2k^2+\beta_{\rm A}^2c_{\rm s}^2k_z^2 -
\biggl[\left(\beta_{\rm FM}^2k^2+\beta_{\rm A}^2c_{\rm s}^2k_z^2\right)^2 -  4\beta_{\rm A}^2c_{\rm s}^2k_z^2k^2\biggr]^{1/2}\biggr\}^{1/2},\nonumber\\
\omega_{\rm FM}&\,=\,& \pm\frac{1}{\sqrt{2}}\biggl\{\beta_{\rm FM}^2k^2+\beta_{\rm A}^2c_{\rm s}^2k_z^2 +
\biggl[\left(\beta_{\rm FM}^2k^2+\beta_{\rm A}^2c_{\rm s}^2k_z^2\right)^2 - 4\beta_{\rm A}^2c_{\rm s}^2k_z^2k^2\biggr]^{1/2}\biggr\}^{1/2}\nonumber\\
&&\label{eq:omega}
\end{eqnarray}
which take the same functional form as in non-relativistic MHD. The
velocity $\beta_{\rm FM}$ is defined by: $\beta_{\rm
  FM}\,\equiv\,\left(\beta_{\rm A}^2+c_{\rm s}^2-\beta_{\rm A}^2c_{\rm
  s}^2\right)^{1/2}$. For convenience, we also define
$c_+\,\equiv\,\left(\beta_{\rm A}^2+c_{\rm s}^2\right)^{1/2}$.

Fast magnetosonic waves thus propagate at a phase velocity
$\vert\omega_{\rm FM}/k\vert$, which ranges from ${\rm
  max}\left(\beta_{\rm A},c_{\rm s}\right)$ at $k_x\,=\,k_y\,=\,0$ to
$\beta_{\rm FM}$ at $k_z\,=\,0$. Slow magnetosonic waves propagate at
a phase velocity $\vert\omega_{\rm SM}/k\vert$, which ranges from $0$
at $k_z\,=\,0$ to ${\rm min}\left(\beta_{\rm A},c_{\rm s}\right)$ at
$k_\perp\,=\,0$. For further details,
see~\citet{goedbloed2010advanced} or \citet{2008PhPl...15j2103K}.

\begin{widetext}
The various MHD modes can be decomposed over the perturbation
variables as follows:
\begin{eqnarray}
  \boldsymbol{\delta\xi}_{{\rm E},\bsk}&\,=\,&\left\{n,0,0,0,0,0,0,0\right\}\delta\psi_{{\rm E},\bsk}\nonumber\\
  \boldsymbol{\delta\xi}_{{\rm A},\bsk}&\,=\,&\left\{0,0,\frac{k_y\omega_{\rm A}}{k_xk_z},-\frac{\omega_{\rm A}}{k_z},0,-\frac{k_y}{k_x}B_z,B_z,0\right\}\delta\psi_{{\rm A},\bsk}\nonumber\\
  \boldsymbol{\delta\xi}_{{\rm M},\bsk}&\,=\,&\left\{\frac{\omega_{\rm M}^2-\beta_{\rm A}^2k^2}
             {c_{\rm s}^2\left(1-\beta_{\rm A}^2\right)\left(k_x^2+k_y^2\right)}n,
  \frac{\omega_{\rm M}^2-\beta_{\rm A}^2k^2}{k_x^2+k_y^2}W,\frac{\omega_{\rm M}k_x}{k_x^2+k_y^2},
    \frac{\omega_{\rm M}k_y}{k_x^2+k_y^2},\frac{c_{\rm s}^2\omega_{\rm M}k_z}{\omega_{\rm M}^2-c_{\rm s}^2k_z^2},
    -\frac{k_xk_z}{k_x^2+k_y^2}B_z,-\frac{k_yk_z}{k_x^2+k_y^2}B_z,B_z\right\}\delta\psi_{{\rm M},\bsk}\nonumber\\
    &&
\label{eq:eigenvec}
\end{eqnarray}
where the subscript $_{\rm M}$ indicates that this applies separately
for all four magnetosonic modes with corresponding frequency
$\omega_{\rm M}$. For completeness, one should include the ghost mode
with $\boldsymbol{\delta\xi}_{{\rm
    G},\bsk}\,=\,\left\{0,0,0,0,0,k_xB_z,k_yB_z,k_zB_z\right\}\delta\psi_{{\rm G},\bsk}$ and
frequency $\omega_{\rm G}\,=\,0$.
\end{widetext}

Given a set of modes, each possibly carrying a different $\omega_{i}$
and $k_{x,i}$ (but assuming that they all share a same $(k_y,k_z)$ as
in Sec.~\ref{sec:dscatt}), one can use the above decomposition to
relate the set of perturbation variables in configuration space to the
8 characteristics of the linearized MHD system:
\begin{equation}
  \delta\xi_i\,=\,\sum_j\,\int\frac{{\rm d}k_{x,j}{\rm d}k_y{\rm d}k_z}{(2\pi)^3}\,
  e^{-i\omega_{j}t+ik_{x,j}x+i\boldsymbol{k_\perp}\cdot\boldsymbol{x_\perp}}\boldsymbol{\mathcal M}_{ij}\delta\psi_j
\label{eq:Riemann-dec}
\end{equation}
introducing the $8\times8$ matrix $\boldsymbol{\mathcal M}$, whose
columns are given by the 8 brackets of the r.h.s. of
Eq.~(\ref{eq:eigenvec}) including the ghost mode, with possibly
different wavenumbers $k_x$ in each separate mode.
 
\bibliographystyle{aasjournal}

\bibliography{shock}

\end{document}